\documentclass[twocolumn,showpacs,preprintnumbers,amsmath,amssymb]{revtex4}

\usepackage{graphicx}
\usepackage{dcolumn}
\usepackage{bm}

\begin{document}


\title{The magnetic structure of Gd$_2$Ti$_2$O$_7$}

\author{M.I. Brammall}
\author{A.K.R. Briffa}
\author{M.W. Long}
\affiliation{%
School of Physics, Birmingham University, Edgbaston, Birmingham, B15 2TT, 
United Kingdom.
}%

\date{\today}

\begin{abstract}
We attempt to solve the magnetic structure of the gadolinium analogue of 
`spin-ice', using a mixture of experimental and theoretical assumptions.  
The eventual predictions are essentially consistent with both the M\"ossbauer 
and neutron measurements but are unrelated to previous proposals.  We find two 
possible distinct states, one of which is coplanar and the other is fully 
three-dimensional.  We predict that close to the initial transition the 
preferred state is coplanar but that at the lowest temperature the ground-state 
becomes fully three-dimensional.  Unfortunately the energetics are consequently 
complicated. There is a dominant nearest-neighbour Heisenberg interaction but then 
a compromise solution for lifting the final degeneracy resulting from a competition 
between longer-range Heisenberg interactions and direct dipolar 
interactions on similar energy scales.
\end{abstract}

\pacs{03.75.Lm, 39.25.+k, 67.40.-w}
\maketitle

\section{\label{sec:level1}Introduction}

`Spin-ice' is now a well studied experimental system with interesting 
fundamental aspects\cite{1}.  Originally the experimental investigation was 
undertaken to try to shed light on magnetic frustration.  The lattice 
geometry is that of a pyrochlore:  a three-dimensional network of 
corner-sharing tetrahedra.  Although the triangle is geometrically frustrated a 
tetrahedron is much more so, culminating in an extensive degeneracy for 
the antiferromagnetic ground-state manifold for a Heisenberg bonded 
pyrochlore\cite{2}.  `Spin-ice' has the required geometry, but has a strong 
orbital and dipolar character that mean that the interactions are not 
Heisenberg-like and the ground-state is not even locally antiferromagnetic.  
This `failure' has led to the current interest\cite{1}, but the original issue 
of pyrochlore magnetism remained.  Other rare-earth atoms are much more 
Heisenberg-like, with gadolinium being archetypal. It is spherically symmetric 
as a consequence of having a half filled shell and, therefore, can be modeled as 
just a large fairly classical spin.  The analogue `spin-ice' material 
Gd$_2$Ti$_2$O$_7$\cite{3} has now been studied and the original issue of a 
Heisenberg pyrochlore has been investigated, subject to any residual 
dipolar issues.

The magnetic structure determination for Gd$_2$Ti$_2$O$_7$ and 
Gd$_2$Sn$_2$O$_7$ is currently a mess.  The neutron scatterers predict bizarre 
states\cite{4,5}, with different spins ordering with different moments and the 
system making no attempt to minimise the natural interactions. In contrast, the 
M\"ossbauer studies\cite{6} offer a simple picture of all sites equally ordered 
and the additional restriction that the moments lie in a particular plane 
perpendicular to a natural local crystallographic direction.  In this article 
we try to rationalise both types of experiment and achieve a consistent magnetic 
structure.

\begin{figure}
\includegraphics[height=7.2 cm, width=8.4 cm]{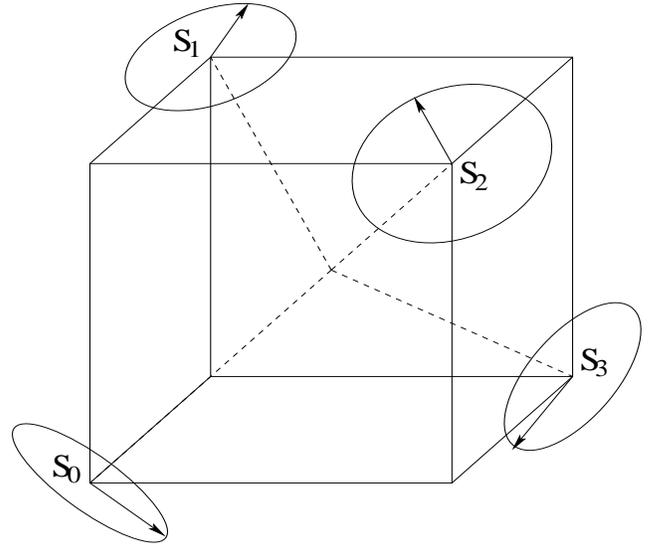}
\caption{\label{fig:1} Restriction of spins to a plane 
perpendicular to the local crystallographic directions pointing to the centre of the 
tetrahedron.}
\end{figure}

The neutron scattering provides a magnetic state indexed by $\Big( \frac{1}{2}, 
\frac{1}{2},\frac{1}{2}\Big) $.  Since the system has an intrinsic four 
atoms per unit cell, this leads to thirty-two atoms per magnetic cell, 
rather more than is comfortable.  We consequently start from the M\"ossbauer 
experiments which suggest that at low temperatures all moments are ordered 
with the restriction that they are ordered perpendicular to the natural local 
crystallographic direction.  (Note that for spin-ice the spins are parallel to 
these local crystallographic directions, which corresponds to a reversal of the 
sign of the anisotropy interaction.)  For a single tetrahedron this restriction is 
depicted in Fig.\ref{fig:1}.
We stress that this restriction is not expected to be a dominant energetic 
restriction because the spin-orbit/dipolar aspects are likely to be very weak 
in comparison to the Heisenberg interactions.  We are employing these 
constraints as experimental restrictions which only become fulfilled to 
extract some extra degeneracy from the choice of ground-state, once all stronger 
Heisenberg interactions have been optimised.  Physically our arguments are 
consequently dangerous since slight avoidance of our constraints can 
provide energetic savings but be lost in the noise of the experiments!

We further enforce an energetic constraint that the total spin in each 
tetrahedron vanishes, which is suggested by the experiment but is not 
guaranteed by it.  This then provides a tractable problem, which annotates the 
residual degeneracy of the Heisenberg model subject to the additional experimentally 
observed single-ion anisotropy.  Note that this antiferromagnetic constraint 
is not applicable to spin-ice\cite{1}, where the orbital and dipolar 
complications provide a very non-Heisenberg interaction.

\section{Magnetic solutions}

The solution of our constrained problem is much simplified using the normalised 
basis: 
\begin{equation}
{\bf \hat S}_0=\left[ \frac{2}{3}\right] ^\frac{1}{2}\left[ {\bf \hat x}\cos 
x_0+{\bf \hat y}\cos y_0+{\bf \hat z}\cos z_0\right] 
\end{equation}
\begin{equation}
{\bf \hat S}_1=\left[ \frac{2}{3}\right] ^\frac{1}{2}\left[ {\bf \hat x}\cos 
x_1-{\bf \hat y}\cos y_1-{\bf \hat z}\cos z_1\right] 
\end{equation}
\begin{equation}
{\bf \hat S}_2=\left[ \frac{2}{3}\right] ^\frac{1}{2}\left[ -{\bf \hat x}\cos  
x_2+{\bf \hat y}\cos y_2-{\bf \hat z}\cos z_2\right] 
\end{equation}
\begin{equation}
{\bf \hat S}_3=\left[ \frac{2}{3}\right] ^\frac{1}{2}\left[ -{\bf \hat x}\cos 
x_3-{\bf \hat y}\cos y_3+{\bf \hat z}\cos z_3\right], 
\end{equation}
which is also subject to 
\begin{equation}
x_\alpha =z_\alpha -\frac{2\pi }{3}\; \; \; \; \; \; \; \; 
y_\alpha =x_\alpha -\frac{2\pi }{3}\; \; \; \; \; \; \; \; 
z_\alpha =y_\alpha -\frac{2\pi }{3}
\end{equation}
all modulo $2\pi $.  This guarantees that the spins lie in the appropriate 
planes and also that they are normalised.  The spins may also be decomposed as: 
\begin{equation}
{\bf \hat S}_0=\left[ \frac{2{\bf \hat x}-{\bf \hat y}-{\bf \hat z}} 
{\surd 6}\right] \cos x_0+\left[ \frac{{\bf \hat y}-{\bf \hat z}}{\surd 2} 
\right] \sin x_0 
\end{equation}
\begin{equation}
{\bf \hat S}_0=\left[ \frac{2{\bf \hat y}-{\bf \hat z}-{\bf \hat x}} 
{\surd 6}\right] \cos y_0+\left[ \frac{{\bf \hat z}-{\bf \hat x}}{\surd 2} 
\right] \sin y_0
\end{equation}
\begin{equation}
{\bf \hat S}_0=\left[ \frac{2{\bf \hat z}-{\bf \hat x}-{\bf \hat y}} 
{\surd 6}\right] \cos z_0+\left[ \frac{{\bf \hat x}-{\bf \hat y}}{\surd 2} 
\right] \sin z_0,
\end{equation}
with analogues for the other three spins, which relates the chosen angles to 
the underlying Cartesian basis.

The solutions to the constraint (see Appendix) 
\begin{equation}
\label{heis}
{\bf \hat S}_0+{\bf \hat S}_1+{\bf \hat S}_2+{\bf \hat S}_3={\bf 0} 
\end{equation}
may be characterised by the different ways of {\it pairing} the atoms.  We have 
one style of solutions with all the angles equal 
\begin{eqnarray}
x_0=x_1=x_2=x_3\; \; \; \; \; \; &&y_0=y_1=y_2=y_3\nonumber \\ 
\; \; \; \; \; \; z_0=z_1=z_2=z_3&&
\end{eqnarray}
and then three associated with pairs 
\begin{eqnarray}
x_0=x_1=-x_2=-x_3\; \; \; \; \; \; &&y_0=-y_1=y_2=-y_3\nonumber \\ 
\; \; \; \; \; \; z_0=-z_1=-z_2=z_3&&
\end{eqnarray}
for the three choices of pairs of pairs.

\begin{figure}
\includegraphics[height=7.2 cm, width=8.4 cm]{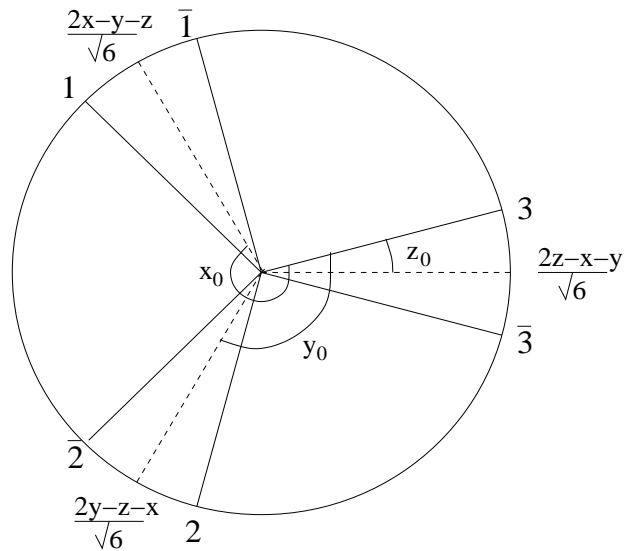}
\caption{\label{fig:2} The six possible angles for a particular 
global spin configuration.  This example is ${\bf S}_0$ and each spin 
configuration is described simultaneously by all three angles, $x_0$, 
$y_0$ and $z_0$.}
\end{figure}

\begin{figure}
\includegraphics[height=7.2 cm, width=8.4 cm]{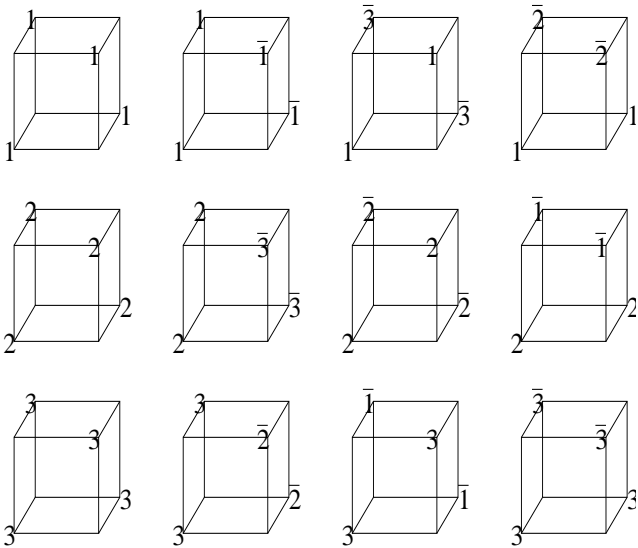}
\caption{\label{fig:3} Possible spin configurations in an 
optimised tetrahedron, using the notation of Fig.\ref{fig:2}.}
\end{figure}

The restrictions to the anisotropy plane and to antiferromagnetism 
severely constrain the permitted magnetic states.  Once we have fixed one 
spin then there are (a maximum of) six permitted orientations for 
each spin in the local basis (a maximum of twenty-four globally).  These 
states are generated by $x_\alpha \mapsto -x_\alpha $, $y_\alpha \mapsto 
-y_\alpha $ and $z_\alpha \mapsto -z_\alpha $, which are the only possibilities 
permitted by the Heisenberg model.  The possible states are annotated in 
Fig.\ref{fig:2}.  One can then generate possible 
global states from the initial spin, using one out of the four possible 
configurations in each subsequent tetrahedron, as depicted in Fig.\ref{fig:3}, 
together with the analogues with $n\mapsto \bar n$ and independently the 
inverted tetrahedra.

The next step is to enumerate the possible global configurations. This 
proves quite difficult, however, so first we set the scene.  For spin-ice the spin 
orientations are quite different, being controlled by `two in and two out' 
for each tetrahedron.  The spins are oriented towards the centres of 
the tetrahedra which induces less technical problems.  All possible 
ground-states may be generated by propagation parallel to a Cartesian 
direction.  One can place an arbitrary collection of spins in a 
particular plane of the geometry which places two spins in each participating 
tetrahedron.  Any pairs that are either both in or both out propagate 
uniquely and each pair that are one in and one out provide two possible 
continuations.  For each possible state we then propagate to the next plane 
generating an exponentially large number of possible ground-states.  Note 
that this is not a solution to the spin-ice-like anisotropy projected 
Heisenberg model, which finds all spins in or all spins out and hence only two 
distinct ground-states because when one spin is chosen then all others are 
constrained.

For our model we follow a two stage process.  Firstly we ignore the `bar' and 
look only at the integer $n\in \{ 1,2,3\} $.  For the special case of $z_0$=0 
this distinction becomes moot anyway.  Indeed, if we think a bit more 
physically, then we would not expect the spins to employ arbitrary directions 
and the special cases of $z_0\in \{ 0,\frac{2\pi }{3},\frac{4\pi }{3}\} $ 
and $z_0\in \{ -\frac{\pi }{6},\frac{\pi }{2},\frac{\pi }{6}\} $ offer 
distinct extra symmetries with both involving only twelve global spin 
orientations.  The first provides a restriction to only three local orientations 
and the second finds commonality between distinct sublattice configurations 
leading to the orientations: 
\begin{equation}
\{ \frac{\pm {\bf \hat y}\pm {\bf \hat z}}{\surd 2},
\frac{\pm {\bf \hat z}\pm {\bf \hat x}}{\surd 2},
\frac{\pm {\bf \hat x}\pm {\bf \hat y}}{\surd 2}\}. 
\end{equation}
This second option is probably the best candidate for the experimental systems.  
By employing propagation along one of the Cartesian directions, we can use a similar 
argument to the previous to investigate the degeneracy.  We focus on a plane 
perpendicular to the $z$-direction for which Fig.\ref{fig:3} provides the 
restrictions that $1$ and $2$ may not be neighbours in the plane but pairs of 
$1$ and $2$ may propagate to the next plane using either $1$ or $2$ for 
the next pair.  This provides a large but seemingly not exponential 
degeneracy, since as well as the new configurations from the choices of $1$ and 
$2$, there is also an associated loss from discounting states where $1$ and $2$ 
are neighbours in the next plane.  The inclusion of the bar further complicates 
matters.

\begin{figure}
\includegraphics[height=1.8 cm, width=8.4 cm]{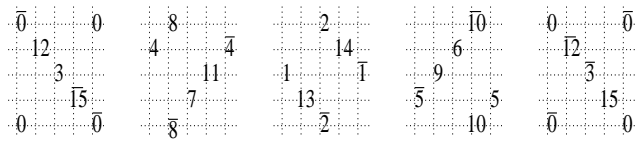}
\caption{\label{fig:4} Distinct atoms in the magnetic unit 
cell.  $\bar n$ are in the reverse direction to $n$.}
\end{figure}

\begin{figure}
\includegraphics[height=4.0 cm, width=8.4 cm]{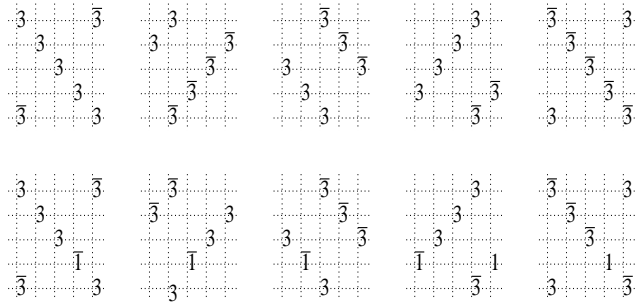}
\caption{\label{fig:5} Possible spin configurations for the 
case $z_0$=$\frac{\pi }{2}$.}
\end{figure}

We now consider the scattering experiments and include the idea that the 
scattering can be indexed using $\Big( \frac{1}{2},\frac{1}{2},\frac{1}{2}\Big) 
$.  At the simplest level this restriction amounts to choosing the sixteen 
spins of Fig.\ref{fig:4}, subject to the previous restrictions of 
Fig.\ref{fig:3}, combined 
with the constraint that the `barred' spins are reversed in orientation.  
In our modeling, this restriction to having reversed spins is also a strong 
constraint.  The easiest way to enforce it is to restrict attention to 
$z_0$=$\frac{\pi }{2}$ and then in Fig.\ref{fig:2} the two spins $n$ and $\bar 
n$ are actually in opposite physical directions.  The previous problem of 
annotating permissible configurations in terms of the states $n$ now provides 
an elementary solution:  we find four styles.  Firstly, pure 
solutions where each site has the same $n$.  Secondly, alternating solutions 
where in a particular direction planes alternate between two distinct values of 
$n$.  Thirdly, tetrahedral solutions with alternating tetrahedra alternating 
between two distinct values of $n$.  Fourthly, period-four solutions where 
one plane in four has a distinct value of $n$.  Fig.\ref{fig:5} shows the only 
two examples which are are consistent with $\Big( \frac{1}{2},\frac{1}{2},
\frac{1}{2}\Big) $. 

\begin{figure}
\includegraphics[height=8.0 cm, width=8.4 cm]{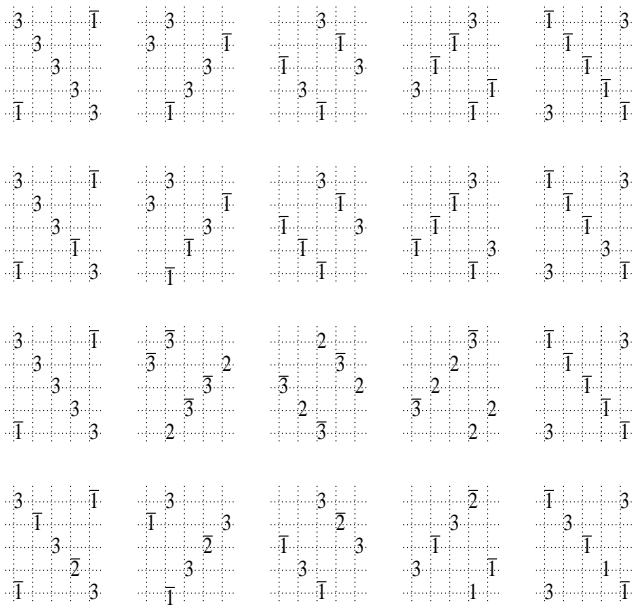}
\caption{\label{fig:6} Possible spin configurations for the 
case $z_0$=-$\frac{\pi }{6}$.}
\end{figure}

A second set of solutions may be obtained from $z_0$=-$\frac{\pi }{6}$ where 
now the pairs $\{ 1,\bar 2\} $ $\{ 2,\bar 3\} $ and $\{ 3,\bar 1\} $ correspond 
to reversed spins.  The permitted solutions are depicted in Fig.\ref{fig:6}, 
and there are only four physically distinct styles.  Clearly by employing 
symmetry there are a mass of rotation and translationally associated analogues 
to the depicted states. In total, however, there appear to be only six styles 
of solutions.  Careful investigation of these states demonstrate that in fact 
there are only two classes unrelated by symmetry, which we can choose to 
be the $z_0$=$\frac{\pi }{2}$ states of Fig.\ref{fig:5}. The first is coplanar 
and the second fully three-dimensional.

\section{Experimental and theoretical issues}

There is one critical final piece in the puzzle:  the form-factors of the 
magnetic Bragg spots.  The internal structure of our proposed states can be 
probed by measuring the relative intensities of the different magnetic Bragg 
spots.  In our notation the strengths of the nearest Bragg spots to the origin 
are directly related to the magnetic moments on each of the four underlying 
sublattices: 
\begin{equation}
{\bf B}_0={\bf S}_0+{\bf S}_1+{\bf S}_2+{\bf S}_3\; \; \; \; \; \; \; 
{\bf B}_1={\bf S}_4+{\bf S}_5+{\bf S}_6+{\bf S}_7
\end{equation}
\begin{equation}
{\bf B}_2={\bf S}_8+{\bf S}_9+{\bf S}_{10}+{\bf S}_{11}\; \; \; \; \; 
\; \; {\bf B}_3={\bf S}_{12}+{\bf S}_{13}+{\bf S}_{14}+{\bf S}_{15}\nonumber. 
\end{equation}
In all of our states these quantities vanish!

\begin{figure}
\includegraphics[height=7.0 cm, width=8.4 cm]{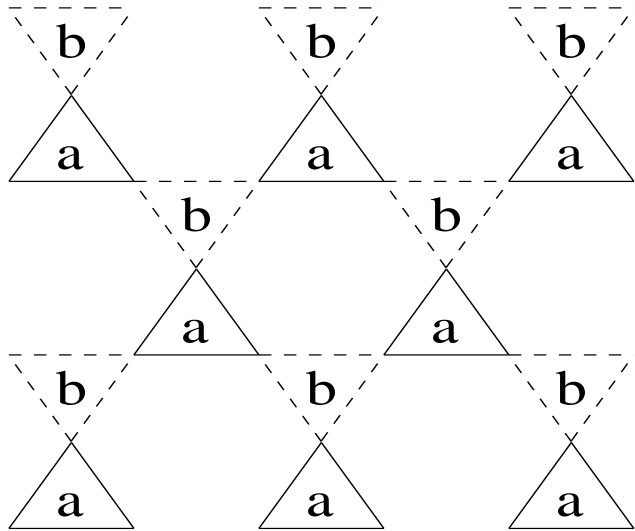}
\caption{\label{fig:7} Kagom\'e plane decomposed into the two 
inversion related triangles.  The (a) denote atoms that complete tetrahedra 
above and the (b) denote atoms that complete tetrahedra below.}
\end{figure}

The actual $\Big( \frac{1}{2},\frac{1}{2},\frac{1}{2}\Big) $ Bragg spot, and 
its symmetrically related neighbours, correspond to decomposing the lattice 
into alternating planes of Kagom\'e connectivity and sparse triangular planes, 
keeping the phase within such a plane uniform and alternating the phase 
between neighbouring equivalent geometries.  Interestingly, our 
antiferromagnetic ansatz directly controls this character.  The Kagom\'e 
planes may be decomposed into two inversion related sets of triangles, as 
depicted in Fig.\ref{fig:7}.  One set (a) corresponds to a set of triangular faces 
of tetrahedra which are completed above the plane, and the other (b) form a 
set of corresponding tetrahedra which are completed below the plane.  
These completing atoms form the neighbouring sparse triangular planes.  The 
imposed constraint that each tetrahedron has zero total spin then 
enforces that the total spin in a Kagom\'e plane is antiparallel to both 
neighbouring triangular planes.  The phase relationship that `neighbouring' 
triangular planes should be antiparallel then requires that the total spin in 
each Kagom\'e plane must vanish.  Consequently the associated Bragg spots must 
also vanish.  Physically the argument only depends on the nearest-neighbour 
Heisenberg interactions:  solving each tertrahedron gives the requirement 
that that tetrahedron should have vanishing total spin.  The Kagom\'e argument 
of Fig.\ref{fig:7} requires that the total spin of each neighbouring 
plane along the (1,1,1) direction (or analogues) must be antiparallel.  Since 
the relevant Bragg spots add every second plane with an alternating sign, the 
appearance of such a Bragg spot is inconsistent with the 
nearest-neighbour Heisenberg interaction.  We have also imposed the additional 
constraint that our solutions have the observed magnetic periodicity which 
requires that the total-spin of each plane vanishes.

This argument provides a second important physical consequence.  In 
terms of the original four atoms per unit cell of the underlying pyrochlore 
structure, the triangular lattices between Kagom\'e planes provide each of the 
four sublattices as the orientation of the Kagom\'e planes is varied.  This fact 
tells us that the total spin of each sublattice independently vanishes.  
Although we have developed our argument employing spin anisotropy and 
experimental periodicity, our chosen states automatically minimise spin 
interactions that we did not impose.  To see this we next analyse third-nearest 
neighbour Heisenberg interactions, which in this case are expected to be 
stronger than the second-nearest neighbour interactions due to the nature of
relavent interaction pathways.  We find that the third-nearest neighbour interactions
are restricted to connect atoms on the same original sublattice.  There are twelve such 
third-neighbours to each spin and they come in two classes.  Firstly, there are six 
neighbours with a magnetic gadolinium atom lying exactly in the middle between 
them.  Secondly, the other six neighbours have a non-magnetic titanium atom lying 
between them.  The gadolinium chains form the original face-centre-cubic 
lattice but with bonds omitted between neighbouring atoms lying in planes 
perpendicular to the appropriate (1,1,1) direction.  In our notation, this interaction 
is optimised by {\it maximising} the total spin of the four atoms on the same 
sublattice rather than minimising it.  The chains of alternating gadolinium and 
titanium atoms yield the sparse triangular lattices in-between the Kagom\'e planes.  
Although our vanishing total spin on these planes is a low energy state for 
these third neighbour interactions, it is only the ground-state of the triangular 
lattice when there are additional longer-neighbour interactions also present.    

We now embark upon a low level analysis of the plausible energetic 
interactions responsible for our proposed states.  There is a sizable 
literature which we do not do justice to.  The most relevant mean-field 
analysis of the Heisenberg interactions is not compatible with the actual 
interactions in this system\cite{7}, using physical distance instead of active 
pathways to choose the relevant interactions.  The dipolar investigations 
are more relevant\cite{3,8} and indicate that the lowest energy solution 
should be a {\bf q}={\bf 0} state\cite{8} although (q,q,q) states should also be 
low energy states\cite{3} and therefore competitive.  We employ much simpler arguments 
to try to get to grips with the most likely physical explanation for the 
observed experimental states.

\begin{figure}
\includegraphics[height=7.2 cm, width=8.4 cm]{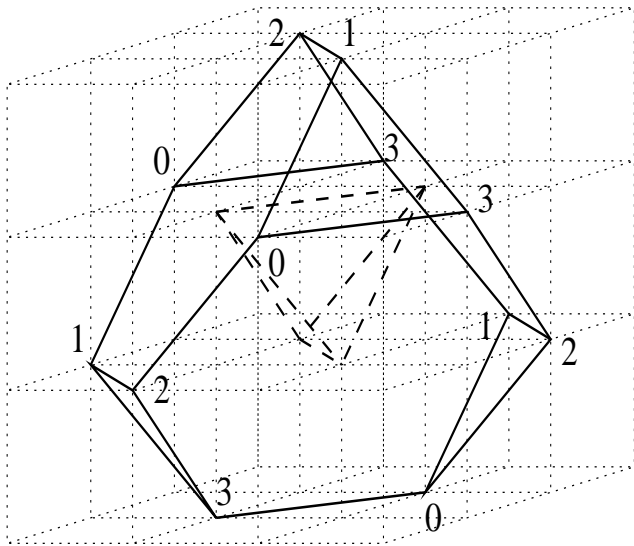}
\caption{\label{fig:8} Cage of gadolinium atoms surrounding a tetrahedron 
of titanium atoms.}
\end{figure}

\begin{figure}
\includegraphics[height=7.2 cm, width=8.4 cm]{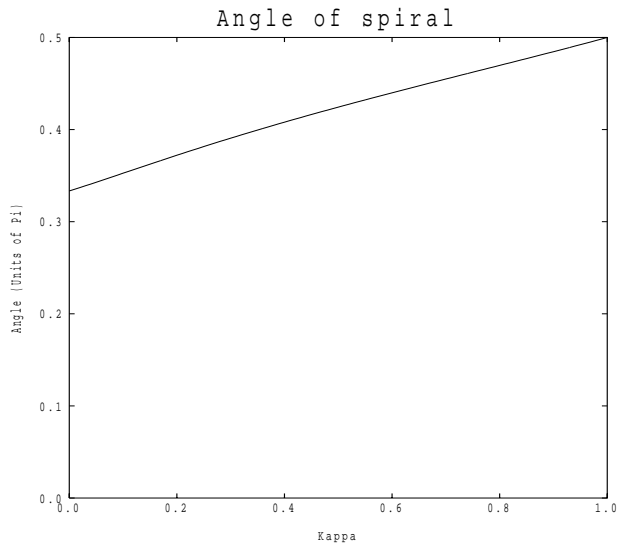}
\caption{\label{fig:9} Angle of the spiral measured in units of $\pi $.}
\end{figure}

\begin{figure}
\includegraphics[height=7.2 cm, width=8.4 cm]{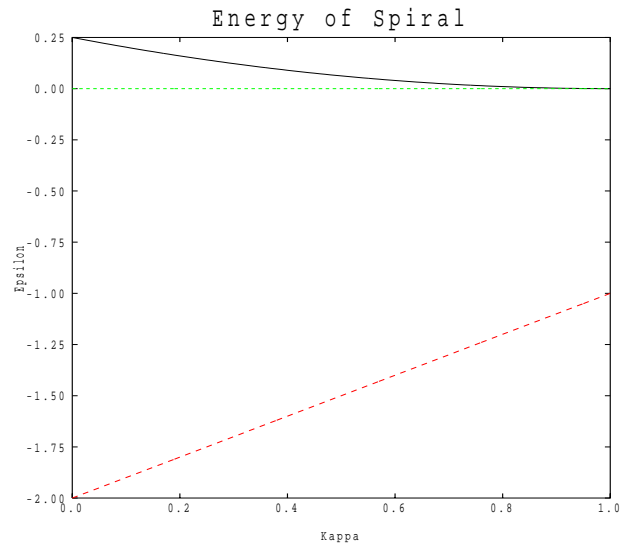}
\caption{\label{fig:10} Energies of various states, ground-state (bold), 
experimentally observed state (dotted) and zone-centre (dashed).}
\end{figure}

The actual additional Heisenberg interactions 
probably stem from pathways across titanium atoms, since the hopping matrix 
elements are so much larger.  The cage of atoms depicted in Fig.\ref{fig:8}
are all expected to develop Heisenberg interactions of various strengths 
between the members, although the bonds between members of the same 
sublattice should be largest.  We employ two new matrix elements for the 
additional Heisenberg interactions across the hexagons depicted in Fig.\ref{fig:8}: 
$\lambda $ for diametric and $\kappa \lambda $ for the remaining two interactions.  
The structure factor is then controlled by the matrix:
\begin{widetext}
\begin{equation}
\Gamma _{\bf k}=\left[ 
\begin{matrix}
\lambda \left( 4xyz-1\right) &X+2\lambda \kappa xc_1&Y+2\lambda \kappa yc_2
&Z+2\lambda \kappa zc_3\\X+2\lambda \kappa xc_1&\lambda \left( 4xYZ-1\right) 
&z+2\lambda \kappa Zc_3&y+2\lambda \kappa Yc_2\\Y+2\lambda \kappa yc_2
&z+2\lambda \kappa Zc_3&\lambda \left( 4XyZ-1\right) &x+2\lambda \kappa Xc_1\\
Z+2\lambda \kappa zc_3&y+2\lambda \kappa Yc_2&x+2\lambda \kappa Xc_1&\lambda 
\left( 4XYz-1\right) \\
\end{matrix}
\right] 
\end{equation}
\end{widetext}
with,
\begin{equation}
x=\cos \frac{k_y-k_z}{2}\; \; y=\cos \frac{k_z-k_x}{2}\; \; z=\cos 
\frac{k_x-k_y}{2}
\end{equation}
\begin{equation}
X=\cos \frac{k_y+k_z}{2}\; \; Y=\cos \frac{k_z+k_x}{2}\; \; Z=\cos 
\frac{k_x+k_y}{2}
\end{equation}
\begin{equation}
c_1=\cos k_x\; \; \; c_2=\cos k_y\; \; \; c_3=\cos k_z
\end{equation}
as the associated parameterisation.  In the absence of the perturbations this 
structure factor has two degenerate bands at $\Gamma = -1$ which control the 
spin degeneracy.  We treat the additional perturbations as small and solve 
for the lifting of this degeneracy in the two degenerate bands.  Employing 
the parameter $\epsilon $ to control the eigenvalues:
\begin{equation}
\Gamma =-1-\lambda -2\lambda \epsilon, 
\end{equation}
we find that the perturbed structure factor satisfies:
\begin{widetext}
\begin{equation}
(1-a_2)\epsilon ^2-2\big( a_2[1+4\kappa ]+3a_2^2-a_1[3a_1+a_3][1+\kappa ]\big) 
\epsilon +3a_2^2[4\kappa ^2-1]-a_3[a_3+2a_1][1+\kappa ]^2+9a_1^2a_2[1-\kappa ^2]
+6a_1a_2a_3[1+\kappa ]-9a_2^3=0
\end{equation}
\end{widetext}
in terms of
\begin{equation}
a_1=\frac{c_1+c_2+c_3}{3}\; \; \; a_2=\frac{c_2c_3+c_3c_1+c_1c_2}{3}\; \; \; 
a_3=c_1c_2c_3.
\end{equation}
This can then be minimised to predict the likely positions of the Bragg spots 
stemming from these physical interactions.  Although the observed experimental 
Bragg scattering is at low energy, there is always a lower energy solution 
that finds a spiral parallel to one of the Cartesian axes.  The ground-state 
as a function of $\kappa $ involves a solution of the form:
\begin{equation}
c_1=c_2=-\cos \theta \; \; \; c_3=1.
\end{equation}
This spiral starts out with $\theta = \frac{\pi }{3}$ when $\kappa = 0$ 
and ends up at the experimental $\theta = \frac{\pi }{2}$ with the unphysical 
value of $\kappa = 1$, as depicted is in Fig.\ref{fig:9}.  
The energy of this ground-state is compared to that of the experimental 
solutions  and the zone-centre solution in Fig.\ref{fig:10}.
It is clear that the second-neighbour Heisenberg interactions are not 
sufficient to describe the physics, and that the zone-centre state has a very 
poor energy.  It therefore appears necessary to resort to the dipolar interactions 
to energetically explain the observed phases.

\begin{figure}
\includegraphics[height=2.2 cm, width=8.4 cm]{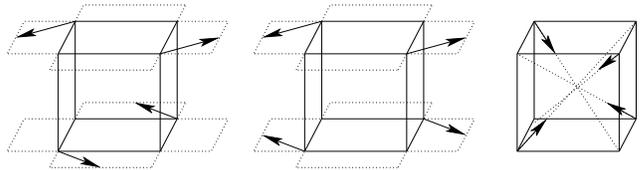}
\caption{\label{fig:11} Total-spin zero eigenstates of the short-range dipolar 
model, in order of energy.}
\end{figure}

Dipolar interactions take the generic form
\begin{equation}
\hat H=\frac{{\bf S}_\alpha .{\bf S}_\beta -3{\bf S}_\alpha .{\bf \hat 
r}_{\alpha \beta }{\bf \hat r}_{\alpha \beta }.{\bf S}_\beta }{\mid 
{\bf r}_{\alpha \beta }\mid ^3}
\end{equation}
where ${\bf r}_{\alpha \beta }\equiv {\bf r}_\alpha -{\bf r}_\beta $ is the 
vector connecting the two interacting spins.  This interaction is 
mathematically taxing because it is both long-range and it relates the 
spin orientation to the lattice directions which breaks spin isotropy.  We will 
extract the initial isotropic Heisenberg interactions which renormalise the 
existing exchange-based interactions and focus only on the second term. This 
second term we further restrict to nearest-neighbours:
\begin{equation}
\label{dipole}
\hat H=-3\sum _{\alpha >\beta }{\bf S}_\alpha .{\bf \hat r}_{\alpha 
\beta }{\bf \hat r}_{\alpha \beta }.{\bf S}_\beta .
\end{equation}
It is elementary to solve this rescaled dipolar interaction on a single 
tetrahedron and we find five styles of solution at energies: 
$\{ -\frac{3}{2}(1+\surd 17),-3,3,\frac{3}{2}(\surd17-1),12\} $.  The 
ground-state has 86.4\% of its moments parallel and is consequently a high 
energy state if the nearest-neighbour Heisenberg interaction dominates.  The 
next state is triply degenerate and as it has zero total-spin it is compatible 
with the Heisenberg interaction.  The third state is doubly degenerate and also 
has zero total-spin.  The fourth state is not compatible as it has 13.6\% of 
its moments parallel.  Finally the fifth state also has zero total-spin.  
The three compatible styles of states are depicted in 
Fig.\ref{fig:11} 
and the crucial observation is that the two low energy states involve spins 
which are perpendicular to the natural crystallographic directions (i.e. the
spins are in the planes indicated by the M\"ossbauer experiments).

The planar spin restrictions deduced from the M\"ossbauer experiments can be 
derived from the nearest-neighbour dipolar interactions subjected to dominant 
nearest-neighbour Heisenberg interactions.  Enforcing equal length spins and 
including the Heisenberg constraints, Eq.(\ref{heis}), we may rewrite the 
dipolar interactions, Eq.(\ref{dipole}), as
\begin{eqnarray}
\hat H=\frac{3}{4}\big[ (S_0^x+S_0^y+S_0^z)^2+(S_1^x-S_1^y-S_1^z)^2
\; \; \; \; \; \; &&\nonumber \\
+(-S_2^x+S_2^y-S_2^z)^2+(-S_3^x-S_3^y+S_3^z)^2\big] &&\nonumber \\
+\frac{3}{8}\big[ (S_0^x+S_1^x-S_2^x-S_3^x)^2
\; \; \; \; \; \; \; \; \; \; \; \; \; \; \; \; \; \; \; \; \; \; \; \; 
&&\nonumber \\
+(S_0^y-S_1^y+S_2^y-S_3^y)^2
\; \; \; \; \; \; \; \; \; \; \; \; \; \; \; \; \; \; \; \; &&\nonumber \\
+(S_0^z-S_1^z-S_2^z+S_3^z)^2\big] 
\; .\; \; \; \; \; \; \; \; \; \; \; \; \; &&
\end{eqnarray}
This is clearly minimised by zeroing each of the seven quadratics, which is 
consistent and indeed uniquely specifies the three states shown later in 
Fig.~\ref{fig:15}.  In particular we note that the first four quadratics may 
be represented by
\begin{eqnarray}
\left[ {\bf S}_0.(1,1,1)\right) ] ^2\; \; \; \; \; \; \; \; \; \; 
\left[ {\bf S}_1.(1,-1,-1)\right] ^2\; \; \; &&\nonumber \\
\left[ {\bf S}_2.(-1,1,-1)\right] ^2\; \; \; \; \; \; \; \; 
\left[ {\bf S}_3.(-1,-1,1)\right] ^2\; .&&
\end{eqnarray}
Orienting the four spins perpendicular to their local crystallographic 
directions therefore partially minimises the local dipolar energy.

One might now conclude that the problem is solved.  We have a dominant 
nearest-neighbour Heisenberg interaction with a weaker dipolar interaction 
which stabilises the observed experimental states, but this is not the case.  
There is a clear solution to the problem of dominant nearest-neighbour 
Heisenberg and weak nearest-neighbour dipolar forces and it is not the 
experimentally observed state.  Instead all that is required is to repeat the 
spiral solution of Fig.\ref{fig:11} in all directions to provide a compatible 
solution that has ${\bf k}={\bf 0}$\cite{8}.  The only way to now rationalise the 
experiments is to assume that there is a {\it competition} between the 
dipolar forces and the longer-range Heisenberg interactions.  The energy 
of the Heisenberg interactions at ${\bf k}={\bf 0}$ is $\epsilon $=-2+$\kappa $ 
which is much worse than $\epsilon $=0 for the experimental solutions.  We can 
then argue that dipolar interactions are not strong enough to overcome 
this energy but are strong enough to select the observed ground-state over 
the weak preference by the second and third-neighbour Heisenberg interactions.

The next task is to assess the dipolar energies in the language that we 
have used to describe our solutions.  The dipolar states which 
do not respect zero total-spin can be split into pieces which are parallel and 
perpendicular to the natural crystallographic directions.  We find the `two in 
and two out' state of spin-ice parallel to the crystallographic directions 
and a saturated ferromagnetic state parallel to a Cartesian direction.  
These states are consequently totally irrelevant.  The high energy state 
is also restricted to the crystallographic directions and hence we are 
restricted to the first two states depicted in Fig.\ref{fig:11}.  Careful 
calculation of the dipolar energies of the six bonds in a tetrahedron then 
yields
\begin{eqnarray}
D=\cos x_0\cos x_1+\cos x_2\cos x_3\; \; &&\nonumber \\
+\cos y_0\cos y_2+\cos y_1\cos y_3\; \; &&\\
+\cos z_0\cos z_3+\cos z_1\cos z_2\; \; &&\nonumber 
\end{eqnarray}
and when we further incorporate our solutions to nearest-neighbour Heisenberg 
interactions we find
\begin{eqnarray}
D_0=3\; \; \; \; \; \; \; \; \; \; \; D_1=3\cos (2x_\alpha )&&\nonumber \\
D_2=3\cos (2y_\alpha )\; \; \; D_3=3\cos (2z_\alpha )&&
\end{eqnarray}
for each of the configurations depicted in the columns of 
Fig.\ref{fig:3} (where $\alpha \in \{ 0,1,2,3\} $ is any of the spins).  
Minimising these energies clearly leads to two of the general dipolar eigenstates 
(shown in Fig. \ref{fig:11}). However, it also highlights the solutions 
$z_0\in \{ \frac{\pi }{2},\pm \frac{\pi }{6}\} $ that we previously required, 
in order to be consistent with $\Big( \frac{1}{2},\frac{1}{2},\frac{1}{2}\Big) $.  
We can now, therefore, choose the value of $z_0$ to optimise the dipolar 
interactions rather than to fit the experiments.  

The final task in this section is to analyse the solutions that we previously 
found for their dipolar content.  The first half of the states depicted in 
both Fig.\ref{fig:5} and Fig.\ref{fig:6} involve half of the tetrahedra being 
in the dipolar ground-state and the other half being in the high energy state.  
The second set of states also involve half of the tetrahedra being in the 
dipolar ground-state but the other half are not in eigenstates of the dipolar 
interaction at all, but instead are superimposed three-quarters the high 
energy state and a quarter the ground-state.  Consequently the more exotic 
states are actually lower in dipolar energy and therefore expected to be favoured 
at low temperatures.

One crucial observation is that for our coplanar states the dipolar 
interactions are energetically consistent with the nearest-neighbour 
Heisenberg interactions while for the non-coplanar states they are not.  
Although the spin state in each tetrahedron is not always the ground-state, 
for the coplanar state it is always a local eigenstate.  For the 
non-coplanar states, however, half of the tetrahedra have states which are 
not local eigenstates. Consequently the system would be expected to  
relax slightly to improve the dipolar energy a little at the expense of the 
nearest-neighbour Heisenberg energy.  This provides a physical mechanism for 
the magnetic Bragg spots closest to the origin to appear!

\section{State characterisation}

Our final task is to try to characterise our different solutions so that they 
might be separated both theoretically and experimentally.  The joint issues 
of quantum fluctuations, thermal disorder and static disorder will all lift 
degeneracy and prefer particular states.  Quantum fluctuations prefer 
collinear states\cite{9}, which allows more coherent fluctuations.  Static 
disorder, however, prefers non-collinear states\cite{10}, which allow static 
distortions in perpendicular directions to the disorder field.  Thermal 
fluctuations prefer the highest density of states at low energies, to 
optimise entropy, so called `order from disorder'\cite{11}. This tends to 
offer similar states to those stabilised by quantum fluctuations.  Our 
states come in two classes, the first half of each set of states in 
Fig.\ref{fig:5} and Fig.\ref{fig:6} are coplanar and the second half involve 
fully three-dimensional spin orientations (although admittedly only eight of 
the twelve conceivable).  We might anticipate the coplanar solutions to be 
stable at low temperature, on quantum fluctuation grounds, but alternatively we 
might expect the non-coplanar states on dipolar grounds.  It is natural to 
associate the observed transition in Gd$_2$Ti$_2$O$_7$ as a transition between 
the two possible states, but we have no overwhelming experimental insights 
that provide clues as to which order of states one might like to pick.

The experiments do provide an important anomaly.  Although in the first 
investigation\cite{4} the Bragg spots closest to the origin were seen to 
vanish, the more careful investigation\cite{5} found small intensities 
which were attributed to the low temperature phase.  As we have pointed out, 
such spots are inconsistent with the nearest-neighbour Heisenberg model.  
There are various options.  Firstly, the observed scattering might be diffuse 
and associated with disorder.  It is known that diffuse scattering is much 
larger than usual in non-collinear magnets\cite{12}.  Secondly, the magnetic 
state is expected to break the cubic symmetry and hence there is no longer a 
requirement that the Heisenberg interaction is perfect.  Thirdly, the 
scattering might be nuclear in origin and hence be associated with a 
tiny structural distortion (although this appears physically quite unlikely).
The dipolar forces argument suggests that coplanar states yield no Bragg 
intensity but that non-coplanar states yield a small one; this is the only 
clue we have to offer.

\begin{figure}
\includegraphics[height=7.0 cm, width=8.6 cm]{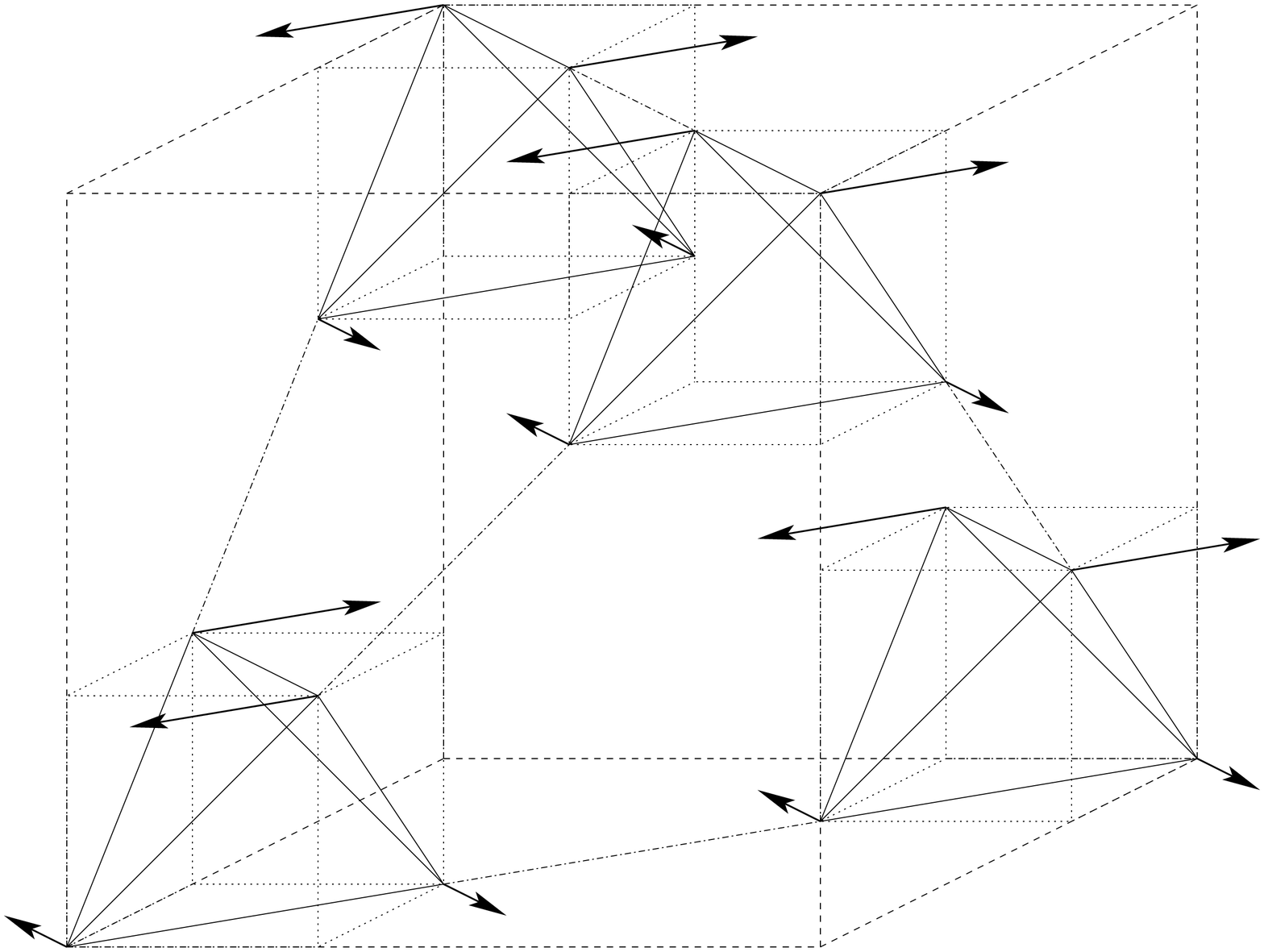}
\caption{\label{fig:12} Coplanar spin arrangement.}
\end{figure}

\begin{figure}
\includegraphics[height=7.0 cm, width=8.6 cm]{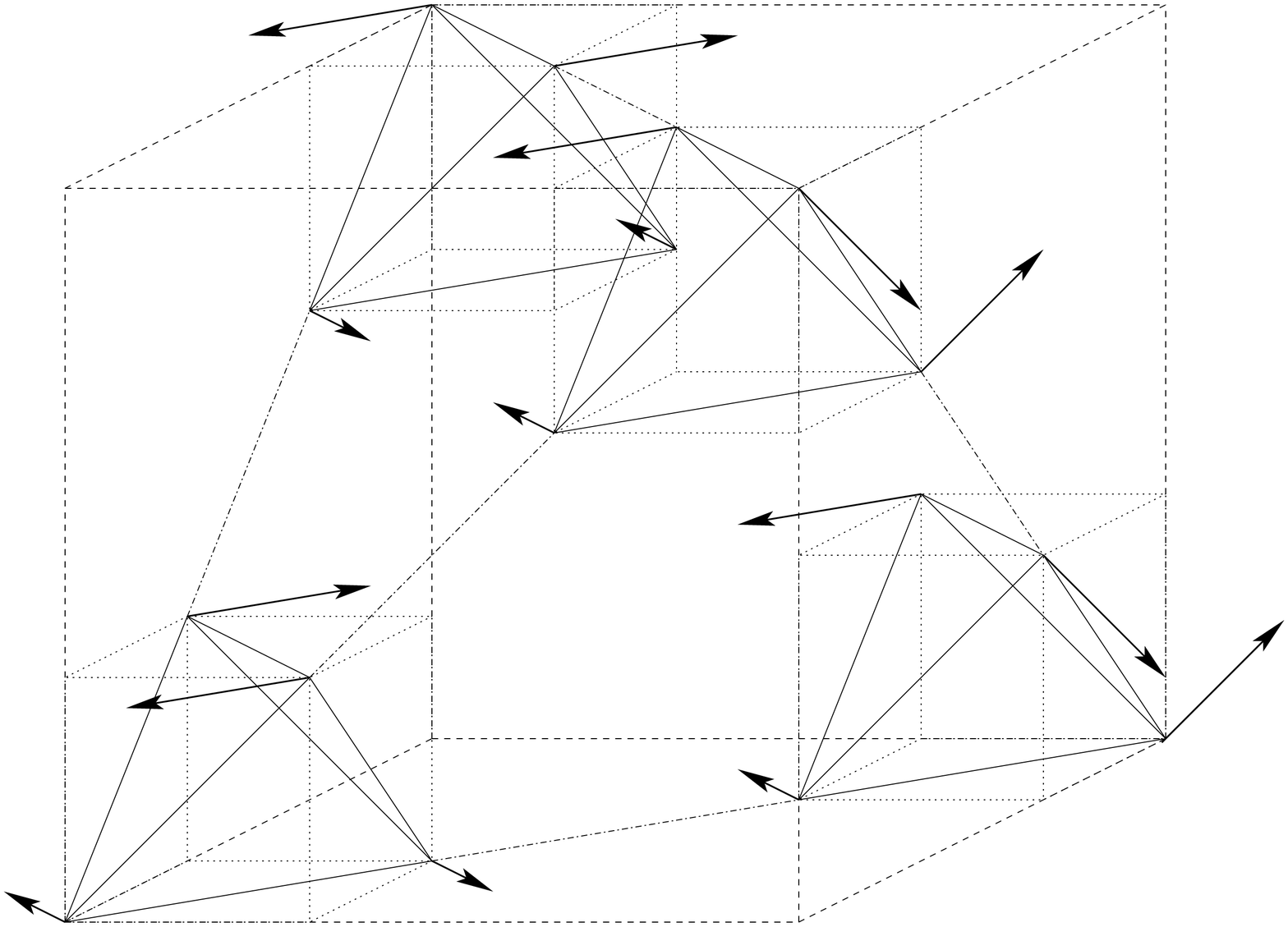}
\caption{\label{fig:13} Non-coplanar spin arrangement.}
\end{figure}

The actual spin arrangements are quite complicated and hence are difficult to 
depict.  We have taken half of the magnetic cell and pictured the coplanar 
and non-coplanar states respectively in Fig.\ref{fig:12} and Fig.\ref{fig:13}. 
The complete spin arrangements are constructed from these by A-B 
ordering a full cubic super-lattice using the depicted cell and its spin 
inverse as the decoration for the super-lattice.

\section{Paradox}

We now arrive at a theoretical paradox: our calculations are 
inconsistent with the current theoretical literature!  We have found a {\it 
complete} set of solutions to our assumptions and the local dipolar energy 
is much worse than the ${\bf q}={\bf 0}$ solution that is expected from the 
`order from disorder' calculations\cite{8}.  The entire ethos behind that 
calculation was the lifting of a degeneracy that remained when the 
dipolar interaction was included, essentially exactly in the classical 
limit at zero temperature.  In our calculations we ought to have found a 
solution that was degenerate with the ${\bf q}={\bf 0}$ solution.

\begin{figure}
\includegraphics[height=3.6 cm, width=4.0 cm]{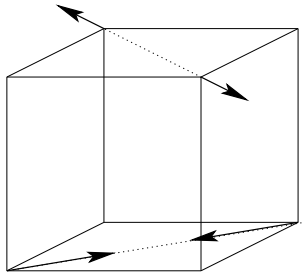}
\caption{\label{fig:14} Possible spin configuration ignored so far.}
\end{figure}

\begin{figure}
\includegraphics[height=3.0 cm, width=8.4 cm]{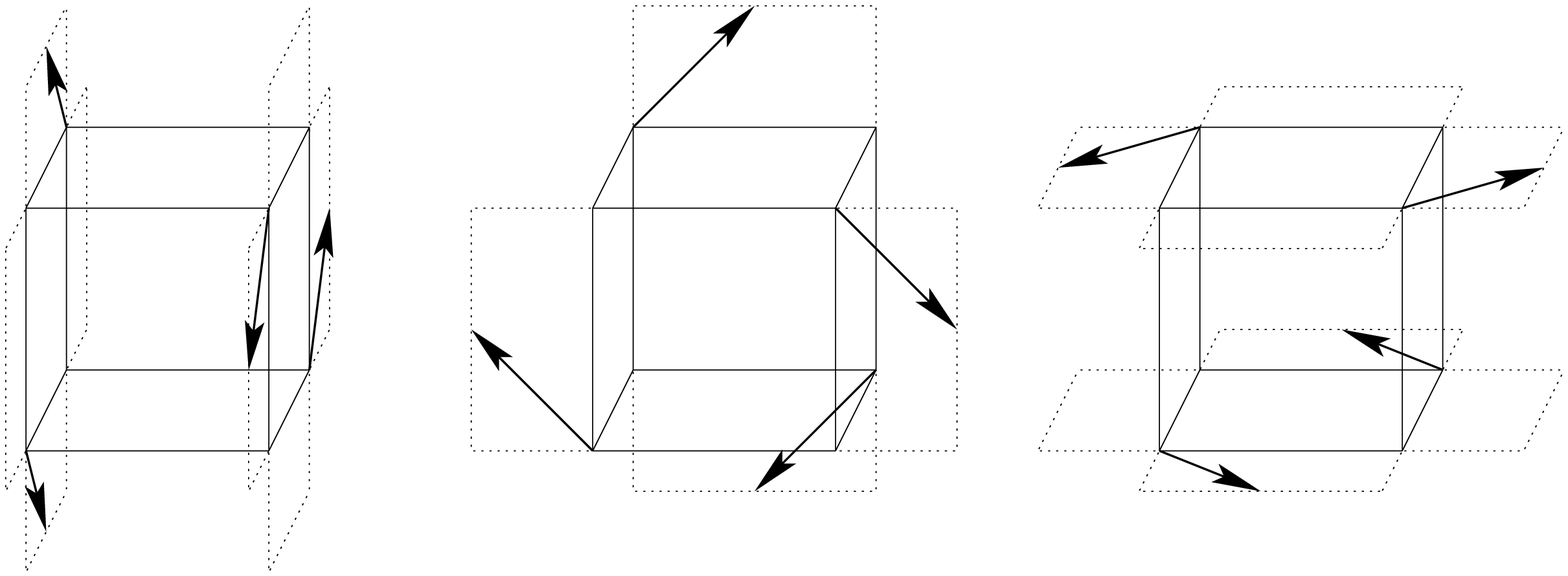}
\caption{\label{fig:15} Local dipolar ground-states when total-spin of 
tetrahedron vanishes.}
\end{figure}

There are various possible resolutions.  Firstly, we have restricted our 
calculations to states which are orthogonal to the natural crystal 
directions and this eliminates two additional zero total-spin states 
from our optimisation.  One is the state with spins pointing to the centre of 
the cube and the other is the state depicted in Fig.\ref{fig:14}. 
Both of these states are of much higher energy than the ground-state.  
Use of these additional states, therefore, cannot lift us to the energy 
of the ${\bf q}={\bf 0}$ solution.  Indeed, we can exactly solve the 
problem of dominant nearest-neighbour Heisenberg and local dipolar interactions: we are 
restricted to the states depicted in Fig.\ref{fig:15}.
These states, which are in the perpendicular subspace, constitute only three 
of the possible configurations in Fig.\ref{fig:3}.  For the case $z_0=0$ they 
are the three configurations with $\{ n,n,\bar n,\bar n\} $, and we can only 
construct {\bf q}={\bf 0} solutions from these configurations  
Secondly, we have only considered the local dipolar interaction and 
the longer-range contributions could overturn our argument.  Although the 
dipolar interaction is long-range, the divergence is irrelevant and we do not 
anticipate issues here.  Thirdly there might be a loop-hole in the previous 
theoretical arguments.  The previous calculations for the Heisenberg and 
dipolar interactions combined\cite{3} calculated the effective structure 
factor for the classical magnetic problem, but they did not proceed on to 
provide an actual magnetic solution.  It was presumed that an appropriate 
magnetic solution would exist.  Unfortunately, this is not guaranteed.  For 
classical magnetism there are additional constraints: the magnitude of the spin of each atom 
on each sublattice must independently be normalised\cite{13}.  For 
Heisenberg-like exchange models there is always an elementary spiral (in all 
bar the Ising model) that will provide a solution.  In this dipolar problem, however, one 
needs to make a multiple-q state involving all four appropriate q's, and there 
are simply not enough spin dimensions  to satisfy all the constraints.  It 
would appear that solving the dipolar problem is harder than first imagined.

\section{More experimental and theoretical issues}

In this section we range across the main experiments that have been performed 
on Gd$_2$Ti$_2$O$_7$ and comment on their relationship to our and previous 
proposals in the literature.  We start out with local probes: M\"ossbauer and 
muon spin rotation.

The M\"ossbauer experiments\cite{6} offer the assertion that the spins are all 
the same length and that they are oriented perpendicular to the local 
crystallographic directions, assumptions that we employed in our modeling.  
More careful reading exhibits the possibility of two magnetic moments 
not both in the preferred planes, but {\it only} in the initial intermediate 
temperature phase.  An additional degree of freedom always provides a 
better fit; the improvement is not significant.  The second magnetic moment 
is also much too large to agree with the states proposed by the neutron scatterers.  
The low temperature phase does not appear to accept a phase with two different 
moments.

The muon spin-rotation experiment\cite{14} offers two clues.  Firstly the 
muon sees two magnetic sites with slightly different fields and a non-magnetic 
site (or a parallel spin).  Secondly there is a sizable relaxation even at 
very low temperatures.  Due to the lack of knowledge about where the muon sits 
(probably in a low symmetry position close to one of the oxygens) the magnetic 
information is not easily usable.  The relaxation tells us that there are 
active low energy excitations disturbing the spins even at the lowest 
temperatures.  Also present in this investigation is some specific heat data, 
which is most instructive.  The majority of the entropy is dissipated above 
the magnetic transitions and the entropy goes to zero at zero temperature 
with a power law.  There is no residual degeneracy at zero temperature 
and the low temperature behaviour is well represented by some low energy 
excitations (surprisingly low energy perhaps).  It is therefore, not possible that 
a macroscopic fraction of the spins are locally disordered in the vicinity of 
the magnetic phase transition.  Any disorder must be highly correlated and 
associated with minimal entropy, viz a few well defined excitations.

There are some very interesting neutron spin-echo measurements\cite{15,16}.  
This pretty experiment gives direct access to the time dependence of the 
magnetic correlations.  The long-time limit clearly shows the expected 
long-range order.  The temporal decay of the measurement exhibits the 
time-scale on which the fluctuations disorder the initial `snapshot' of the 
spin correlations.  Thermal fluctuations are expected to decay at low 
temperature and so the measurement flattens at low temperature\cite{16}.  The 
most intriguing aspect is the fact that on observable time-scales the 
signal does not saturate, but appears to converge to three-quarters!  
This is taken to mean that a quarter of the spins are fluctuating, but from 
the specific heat data, they would have to be doing this coherently.  It is 
not clear from the papers how this normalisation is accomplished, because 
naively one takes a very low temperature and zero field to normalise the data, 
assuming that the system has no residual dynamic fluctuations there.  If 
this result is correct, then it is certainly interesting.

Finally we come to the elastic neutron scattering\cite{4,5}, with the proposal 
of states with spins of different average length.  Although it is stated that 
the neutron scattering promotes only these states, the current paper 
clearly shows that states exist which are consistent with the elastic neutron 
scattering and the M\"ossbauer restrictions that the spins are essentially of 
the same length.  We now consider the theoretical ideas and what they say 
about the proposed states.  The states with spins of different average lengths 
require fluctuations to explain the reduction in lengths, either 
thermal or quantum in nature.  At low temperature the entropy measurements 
indicate that thermal fluctuations are irrelevant and so we are left solely 
with quantum fluctuations.  This is a possible explanation, because 
quantum fluctuations gain energy by making spins locally antiparallel in 
directions perpendicular to the classical order.  They do not require any entropy, 
as they amount to a particular phase for the fluctuation and not a random 
one.  Low dimensional systems are particularly susceptible to quantum 
fluctuations, as are frustrated systems, although larger spins are less 
susceptible as the fractional energy gain is proportional to the inverse of the 
length of the spin.  The situation that we have is abnormal because we 
anticipate that there are two energy scales: the highest wants all tetrahedra 
to have zero total-spin and the second lifts the residual degeneracy and 
promotes the observed long-range order.  A study of a single tetrahedron sets 
the fluctuation scale.  The classical energy of a single terahedron is 
$-2JS^2$ and the quantum energy is $-2JS^2-2JS$.  This means that the quantum fluctuations 
have access to $\frac{2}{7}$ of the classical energy for our system.  
Obviously, a fluctuation in one tetrahedron corrupts two others and so only 
a small fraction of this energy is available to all tetrahedra simultaneously.  
In our proposed states we would expect quantum fluctuations to exist and to 
reduce the observable lengths of the ordered moments.  There is no expectation 
that some spins would fluctuate vastly more than others.  This is a 
plausible interpretation for the lack of saturation in the neutron spin-echo 
experiments.  For the variable-length spin states, we need to assume that the 
quantum fluctuations are strong for some spins and weak for others.  Once 
again, this is not impossible, because the idea of dimerisation can be extended 
to that of independent tetrahedra with quantum states, at a stretch.  The 
problem is that one needs to gain more than one loses.  The states proposed 
by the elastic neutron scatterers involve some spins with essentially saturated 
classical moments and some with tiny moments.  If the moment is saturated then 
it cannot fluctuate.  The spins with tiny moments are either well 
separated on sparse triangular planes, or combined on particular tetrahedra.  
The first possibility is energetically awful, because no coherent fluctuations 
can develop and incoherent fluctuations gain no energy and contradict the 
specific heat data.  The second possibility is only problematic because the 
energetics is poor.  Each fluctuating tetrahedron has access to $-7J$ of 
extra energy, but the four connected tetrahedra have three correlated 
saturated spins in the 120$^\circ $ phase and consequently naturally lose 
$-24.5J$, a very bad deal!  The variable-length spin phases, therefore, are not 
consistent with the theoretical assumption that the dominant energy is the 
nearest-neighbour Heisenberg interaction.

Experiments on Gd$_2$Sn$_2$O$_7$ indicate that the ground-state is actually 
the dipolar preferred ${\bf q}={\bf 0}$ phase\cite{17}.  At first sight this is 
surprising, since the degeneracy is on the gadolinium atom and the tin or 
titanium atom appears passive.  However, we need to employ higher-order exchange 
paths across this `passive' atom and if we assume that the titanium 
$d$-shell is closer to the chemical potential than the tin $p$-shell, then we 
can clearly expect a difference in the longer-range Heisenberg interactions.  This 
would indicate that for the tin compound only nearest-neighbour Heisenberg and dipolar 
interactions are required to explain it.

\section{Conclusion}

We have employed a hybrid method to determine plausible magnetic ground-states 
for the pyrochlore magnet Gd$_2$TI$_2$O$_7$.  We used the theoretical ansatz 
that the nearest-neighbour Heisenberg model is minimised, the experimental 
observations that the magnetic Bragg scattering is indexed by $\big( 
\frac{1}{2},\frac{1}{2},\frac{1}{2}\big) $, and that all spins have 
equivalent moments which are oriented perpendicular to the local natural 
crystallographic directions.  These assumptions provide only two distinct 
classes of solutions and we expect the system to exhibit one or both of these 
phases to explain the observed phase diagram.  Our best guess is that close to 
the initial transition the thermal fluctuations prefer a coplanar state but 
that at low temperature the dipolar interactions energetically stabilise a 
non-coplanar state.  The observed appearance of a very weak intensity to the 
closest magnetic Bragg spots to the origin only in the low temperature phase 
is consistent with this prediction.  We await more detailed experimental form 
factor analysis to confirm or deny this proposal.

The field dependent phase diagram\cite{18} seems particularly instructive, but 
currently is too difficult for us to predict.

\begin{acknowledgments}
We wish to acknowledge useful discussions with A.J. Schofield.
\end{acknowledgments}

\appendix

\section{}

In this appendix we solve the trigonometric constraints 
\begin{equation}
\cos x_0+\cos x_1-\cos x_2-\cos x_3=0
\label{eqx}
\end{equation}
\begin{equation}
\cos y_0-\cos y_1+\cos y_2-\cos y_3=0
\label{eqy}
\end{equation}
\begin{equation}
\cos z_0-\cos z_1-\cos z_2+\cos z_3=0.
\label{eqz}
\end{equation}
Firstly we break the symmetry and focus on $z_\alpha $, rewriting the first two 
equations provides 
\begin{eqnarray}
\sin \left( \frac{z_0-z_3}{2}\right) \sin \left( \frac{z_0+z_3}{2}-\frac{2\pi }
{3}\right) &&\nonumber \\+\sin \left( \frac{z_1-z_2}{2}\right) \sin \left( 
\frac{z_1+z_2}{2}-\frac{2\pi }{3}\right) &&=0
\end{eqnarray}
\begin{eqnarray}
\sin \left( \frac{z_0-z_3}{2}\right) \sin \left( \frac{z_0+z_3}{2}+\frac{2\pi }
{3}\right) &&\nonumber \\-\sin \left( \frac{z_1-z_2}{2}\right) \sin \left( 
\frac{z_1+z_2}{2}+\frac{2\pi }{3}\right) &&=0
\end{eqnarray}
and then mixing them offers 
\begin{eqnarray}
\sin \left( \frac{z_0-z_3}{2}\right) \sin \left( \frac{z_0+z_3}{2}\right) 
&&\nonumber \\+\surd 3\sin \left( \frac{z_1-z_2}{2}\right) \cos \left( 
\frac{z_1+z_2}{2}\right) &&=0
\label{mix1}
\end{eqnarray}
\begin{eqnarray}
\surd 3\sin \left( \frac{z_0-z_3}{2}\right) \cos \left( \frac{z_0+z_3}{2}
\right) &&\nonumber \\+\sin \left( \frac{z_1-z_2}{2}\right) \sin \left( 
\frac{z_1+z_2}{2}\right) &&=0.
\label{mix2}
\end{eqnarray}
Including the third original equation as 
\begin{eqnarray}
\cos \left( \frac{z_0-z_3}{2}\right) \cos \left( \frac{z_0+z_3}{2}\right) 
&&\nonumber \\-\cos \left( \frac{z_1-z_2}{2}\right) \cos \left( 
\frac{z_1+z_2}{2}\right) &&=0
\end{eqnarray}
we can now eliminate $z_1$ and $z_2$ to provide 
\begin{equation}
s_-^2\left( 4s_+^2-3\right) \left( \left[ 9-8s_+^2\right] \left[ 1-s_-^2
\right] +2s_+^2\right) =0
\end{equation}
in terms of 
\begin{equation}
s_\pm =\sin \left( \frac{z_0\pm z_3}{2}\right). 
\end{equation}
The final complicated solution is unphysical so we generate two independent 
solutions.  Firstly (1) 
\begin{equation}
\sin \left( \frac{z_0-z_3}{2}\right) =0\; \; \; \; \Rightarrow \; \; \; \; 
\sin \left( \frac{z_1-z_2}{2}\right) =0
\end{equation}
and consequently (modulo $2\pi $) 
\begin{equation}
z_0=z_3\; \; \; \; z_1=z_2\; \; \; \; \; \cos z_0=\cos z_1
\end{equation}
and the two solutions 
\begin{equation}
z_0=z_1=z_2=z_3\; \; \; \; \; \; z_0=-z_1=-z_2=z_3.
\end{equation}
Secondly (2) 
\begin{equation}
\sin ^2\left( \frac{z_0+z_3}{2}\right) =\frac{3}{4}\; \; \; \; \Rightarrow 
\; \; \; \; \cos (z_0+z_3)=-\frac{1}{2}
\end{equation}
and from Eq.\ref{mix1} and Eq.\ref{mix2}
\begin{equation}
\sin \left( \frac{z_1-z_2}{2}\right) \sin \left( \frac{z_0+z_3}{2}-
\frac{z_1+z_2}{2}\right) =0.
\end{equation}
The solution $z_1$=$z_2$ provides a subset of the solutions of (1), and the new 
possibility is 
\begin{equation}
z_0+z_3=z_1+z_2
\end{equation}
leading to two new solutions: (2.i) 
\begin{equation}
z_0+z_3=\frac{4\pi }{3}=z_1+z_2\; \; \; \; \Rightarrow \; \; \; \; 
x_0+x_3=0=x_1+x_2
\end{equation}
which reduces the original equations Eq.\ref{eqy} and Eq.\ref{eqz} to 
\begin{eqnarray}
\cos \left( x_0-\frac{2\pi }{3}\right) -\cos \left( x_1-\frac{2\pi }{3}
\right) &&\nonumber \\+\cos \left( x_1+\frac{2\pi }{3}\right) -\cos \left( 
x_0+\frac{2\pi }{3}\right) &&=0
\end{eqnarray}
\begin{eqnarray}
\cos \left( x_0+\frac{2\pi }{3}\right) -\cos \left( x_1+\frac{2\pi }{3}
\right) &&\nonumber \\-\cos \left( x_1-\frac{2\pi }{3}\right) +\cos \left( x_0-
\frac{2\pi }{3}\right) &&=0
\end{eqnarray}
and the unique solution 
\begin{equation}
x_0=x_1=-x_2=-x_3.
\end{equation}
Alternatively we can have: (2.ii) 
\begin{equation}
z_0+z_3=-\frac{4\pi }{3}=z_1+z_2\; \; \; \; \Rightarrow \; \; \; \; 
y_0+y_3=0=y_1+y_2
\end{equation}
which reduces the original equations Eq.\ref{eqx} and Eq.\ref{eqz} to 
\begin{eqnarray}
\cos \left( y_0+\frac{2\pi }{3}\right) +\cos \left( y_2-\frac{2\pi }{3}
\right) &&\nonumber \\-\cos \left( y_2+\frac{2\pi }{3}\right) -\cos \left( 
y_0-\frac{2\pi }{3}\right) &&=0
\end{eqnarray}
\begin{eqnarray}
\cos \left( y_0-\frac{2\pi }{3}\right) -\cos \left( y_2+\frac{2\pi }{3}
\right) &&\nonumber \\-\cos \left( y_2-\frac{2\pi }{3}\right) +\cos \left( 
y_0+\frac{2\pi }{3}\right) &&=0
\end{eqnarray}
and the unique solution 
\begin{equation}
y_0=-y_1=y_2=-y_3.
\end{equation}


\end{document}